\documentclass[twoside,english,sort&compress]{iopart}
\usepackage[T1]{fontenc}
\usepackage[latin9]{inputenc}
\usepackage{geometry}
\usepackage{float}
\usepackage{amstext}
\usepackage{graphicx}
\usepackage[numbers]{natbib}

\makeatletter
\usepackage{iopams}
\usepackage{setstack}

\newcommand{\eqref}[1]{(\ref{#1})}

\makeatother

\usepackage{babel}
\begin{document}

\title{Abrupt degenerately-doped silicon nanowire tunnel junctions}

\author{{\Large{}Cristina Cordoba$^{1}$, Taylor S Teitsworth$^{2}$, Mingze
Yang$^{1}$, James F Cahoon$^{2}$ and Karen L Kavanagh$^{1}$} }

\address{{\large{}$1$ Department of Physics, Simon Fraser University, 8888
University Drive, Burnaby, British Columbia V5A 1S6, Canada}}

\address{{\large{}$2$ Department of Chemistry, University of North Carolina
at Chapel Hill, Chapel Hill, North Carolina 27599-3290, United States}}

\ead{{\large{}jfcahoon@unc.edu; kavanagh@sfu.ca}}
\begin{abstract}
We have confirmed the presence of narrow, degenerately-doped axial
silicon nanowire (SiNW) \emph{p-n} junctions via off-axis electron
holography (EH). SiNWs were grown via the vapor-solid-liquid (VLS)
mechanism using gold (Au) as the catalyst, silane (SiH$_{4}$), diborane
(B$_{2}$H$_{6}$) and phosphine (PH$_{3}$) as the precursors, and
hydrochloric acid (HCl) to stabilize the growth. Two types of growth
were carried out, and in each case we explored growth with both \emph{n}/\emph{p}
and\emph{ p}/\emph{n} sequences. In the first type, we abruptly switched
the dopant precursors at the desired junction location, and in the
second type we slowed the growth rate at the junction to allow the
dopants to readily leave the Au catalyst. We demonstrate degenerately-doped
\emph{p}/\emph{n} and \emph{n}/\emph{p} nanowire segments with abrupt
potential profiles of $1.02\pm0.02$ and $0.86\pm0.3$ V, and depletion
region widths as narrow as $10\pm1$ nm via EH. Low temperature current-voltage
measurements show an asymmetric curvature in the forward direction
that resemble planar gold-doped tunnel junctions, where the tunneling
current is hidden by a large excess current. The results presented
herein show that the direct VLS growth of degenerately-doped axial
SiNW \emph{p-n} junctions is feasible, an essential step in the fabrication
of more complex SiNW-based devices for electronics and solar energy.
\end{abstract}

\noindent{\it Keywords\/}: {Silicon nanowires, tunnel junctions, off-axis electron holography}

\pacs{11}

\ams{44}

\submitto{\NT }
\maketitle

\section{Introduction}

Silicon (Si) continues to hold a prevalent position at the heart of
solar energy conversion and the microelectronics industry. Its abundance,
non-toxicity and excellent photoelectric behavior have not only made
Si ubiquitous in these fields but have also encouraged the development
of technologies for its extraction and processing, making it very
competitive in terms of clean energy production \citep{SinghR,Huenteler,YangY}.
Silicon nanowires (SiNWs) are especially attractive for a variety
of applications that benefit from the large surface-area-to-volume
ratio of this pseudo 1-dimensional geometry, including sensors \citep{Shehada,Zhang16,DemamiF},
electrodes \citep{Xiaolei,Aradilla17}, and energy harvesters \citep{LaPierre}.

One type of electronic junction that is a challenge to fabricate,
especially for nanowires (NWs) created via vapor-liquid-solid (VLS)
growth mechanisms, is the tunnel (Esaki) diode. In the simplest case,
these are homo-junctions consisting of abrupt transitions between
degenerately-doped \emph{p} and\emph{ n}-type segments of the same
semiconductor. Narrow depletions regions, less than 10 nm are optimal,
allowing carriers to tunnel through the potential energy barrier at
low applied biases \citep{Esaki}. The ideal current-voltage (\emph{I-V})
characteristics show an interesting negative-differential resistance
(NDR) curve. Lattice-mismatched NW heterojunctions, including InP/Si
\citep{Heurlin}, InP/GaAs \citep{WallentinPersson}, InP/GaInP \citep{Xulu}
and GaSb/InAs \citep{Ek,Ganjipour,B.Mattias} have been reported to
form tunnel junctions. The latter exploited a type II band alignment
that can enhance tunneling probabilities without further doping.

Many factors impede the formation of the perfect structure. A catalyst
reservoir effect may broaden the width of the depletion region by
prolonging the release of a dopant precursor after the vapor-phase
sources are turned off and/or switched \citep{Heurlin,WallentinPersson,Christesen}.
The limited solubility of different precursors at characteristic VLS
growth temperatures can hinder the ability to reach degenerate-doping
profiles \citep{SchmidH}. Other challenges are related to catalyst
segregation, unwanted stacking faults, kinking and radial growth,
and the formation of undesired radial junctions in addition to the
axial junction.

SiNW growth using silane (SiH$_{4}$) as the Si source and phosphine
(PH$_{3}$) and diborane (B$_{2}$H$_{6}$) as the \emph{n-} and \emph{p}-type
dopant precursors, respectively, is well established \citep{Hill2017,Kempa,Pinion,Kendrick,Christesen}.
It is possible to suppress the reservoir effect to form very abrupt
degenerately-doped \emph{n}-type-to-intrinsic transitions by choosing
growth conditions in which the P evaporation (associative desorption)
greatly dominates over crystallization at the liquid-solid interface
\citep{Christesen2016,Christesen}. However, growth of degenerately-doped
\emph{p}-type SiNWs (B-doped) has been challenging and has impeded
the creation of a SiNW tunnel junction except where a degenerate \emph{p}-type
doped Si substrate is used\citep{SchmidH}. While higher temperatures
aid the uniformity of radial dopant distributions, this also increases
the decomposition rate of the SiH$_{4}$ precursor, creating a non-selective
vapor-solid (VS) Si shell growth on the NW sidewalls \citep{Pan}.
Diborane also catalyzes the decomposition SiH$_{4}$, intensifying
the \emph{p}-type growth problems \citep{Hill2017}. Sidewall deposition
can be eliminated by adding hydrochloric acid (HCl), which is known
to create monochlorinated surface species that passivate surfaces
and act as a barrier to deposition. Using HCl passivation and lower
growth temperatures, degenerate doping above the thermodynamic solubility
limit has recently been achieved in \emph{p}-type SiNWs with no evident
VS deposition on the sidewalls \citep{Hill2017} allowing for the
growth of \emph{p}-\emph{n} SiNW superlattices. 

Evaluation of the depletion region width in a NW p-n junction can
be carried out through various methods, including wet chemical etching
\citep{Christesen2013}, secondary electron (SE) imaging \citep{Sealy,Molotskii},
off-axis electron holography (EH) \citep{Yazdi,Darbandi,Hertog,Gan1},
electron beam induced current (EBIC) \citep{HAN2017,DarbandiEBIC},
Kelvin probe force microscopy \citep{Koren,Vinaji}, and other AFM-based
techniques such as scanning capacitance microscopy (SCM) \citep{Brouzet,Vallett}
and scattering-type scanning near-field optical microscopy \citep{Ritchie}.
Methods based on secondary ion mass spectroscopy including atom probe
tomography are very powerful at mapping dopants \citep{Sun}, but
they cannot evaluate the level of activation in the dopant impurities
detected. Wet chemical etching, SE imaging and EBIC provide spatial
junction profile information but are unable to evaluate junction potential
directly, and AFM techniques often require additional information
to separate surface from bulk potentials. EH is a transmission electron
microscopy (TEM) technique that provides two dimensional maps of the
average built-in potential $V_{bi}$ of a \emph{p-n} junction with
a resolution $<$ 5 nm. It is becoming increasingly used in the characterization
of NWs, \citep{Wu,Darbandi,Cooper,Yazdi2013} including SiNWs \citep{Gan1,Hertog,He}.

The phase change of an electron wave after passing through a non-magnetic
specimen of thickness $t$ is given by
\begin{equation}
\Delta\phi(x,y)=C_{E}\int_{0}^{t}V(x,y,z)dz,
\end{equation}
for low dynamical scattering conditions, away from strong diffraction.
Here, $C_{E}$ is an electron-energy-dependent interaction constant
($7.29\times10^{6}$ rads $\text{V}^{-1}\text{m}^{-1}$ at 200 keV),
$z$ is the direction of the electron beam, and $V$ is the electrostatic
potential which includes the periodic potential of the crystal, known
as the mean inner potential (MIP), the built-in junction potential
($V_{bi}$) of the space charge region, and other possible potentials
arising from charge accumulation induced by the electron beam. NWs
are usually the perfect candidates for EH, because they are highly
symmetric and require no chemical or mechanical thinning to become
electron transparent. If the NW is grounded to the grid, then the
degenerately-doped junction is assumed to be unbiased and the chemical
potential $\mu$ (i.e. the Fermi level) will be independent of position
across the junction. The potential drop at a degenerately-doped \emph{p-n
}junction is approximately the value of the bandgap.

EH has been used to image lateral $V_{bi}$ gradients in GaP core-shell
NW \emph{p-n} junctions and to assess the efficiency of dopant incorporation
\citep{Yazdi}. It was also used to measure the abruptness of an axial,
degenerately-doped, GaAs NW homojunction finding a depletion width
of $74\pm9$ nm with a junction potential comparable to the bandgap,
$1.5\pm0.1$ eV. A small diameter reduction in the NW occurring before
the \emph{p-n} junction formation also lead to the conclusion that
there was a delay in the diethylzinc (DEZn) incorporation \citep{Darbandi}.
Gan \emph{et} \emph{al.} mapped the electrostatic potential across
axial SiNWs \emph{p-n} junctions through EH, finding a junction height
of $1.0\pm0.3$ V, with a Schottky barrier at the Au metal contact
of $0.5$ \citep{Gan1}. The $V_{bi}$ profiles in SiNWs junctions
have been interpreted to have a doping profile abruptness of less
than 30 nm/decade\citep{Hertog}.

In this paper, SiNWs were grown via HCl-stabilized VLS growth with
Au catalysts onto thermally-oxidized (300 nm) Si (001) wafers. Two
growth sequences, \emph{n}/\emph{p} and \emph{p}/\emph{n}, with degenerate-doping
levels were synthesized. The effect of the precursor gas flow process
and growth rate at the junction regions on SiNW properties were investigated
towards reduction of Au segregation and improvement in dopant incorporation
and abruptness. We used EH to map the $V_{bi}$ and junction depletion
width. Depletion regions as short as $10\pm1$ nm and $V_{bi}$ values
that are comparable to the Si bandgap were measured. Low temperature
\emph{I-V} measurements showed an effective low-bias resistance as
low as $1\times10^{15}$ $\Omega/\text{cm}^{2}$ and a slope change
in the forward-bias direction characteristic of Au-doped \emph{p-n}
tunnel junctions.

\section*{Results}

Figure 1 shows gas flow schemes for the two growth processes that
were compared for each sequence: \emph{n}/\emph{p} and \emph{p}/\emph{n}.
In the standard process (Growth 1), the flow rate of the Si gas flow
(SiH$_{4}$) was unchanged while the dopant-precursor flows were abruptly
changed at the junction location. In the modified growth process (Growth
2), the SiH$_{4}$ flow rate was reduced from the standard rate by
an order of magnitude, to reduce the growth rate, before the junction
region, and then increased back to the original flow rates after the
junction. The flow rate of the dopant precursors, was also reduced
to maintain the same Si:dopant ratio. Growth 2 was designed to test
if slowing the NW growth rate and allowing more time for dopant atoms
to evaporate from the catalyst would facilitate an increase in junction
abruptness \citep{Christesen,Hill2017,Pinion,Kim}.

\begin{figure}[!htb]
\begin{centering}
\includegraphics[scale=1.2]{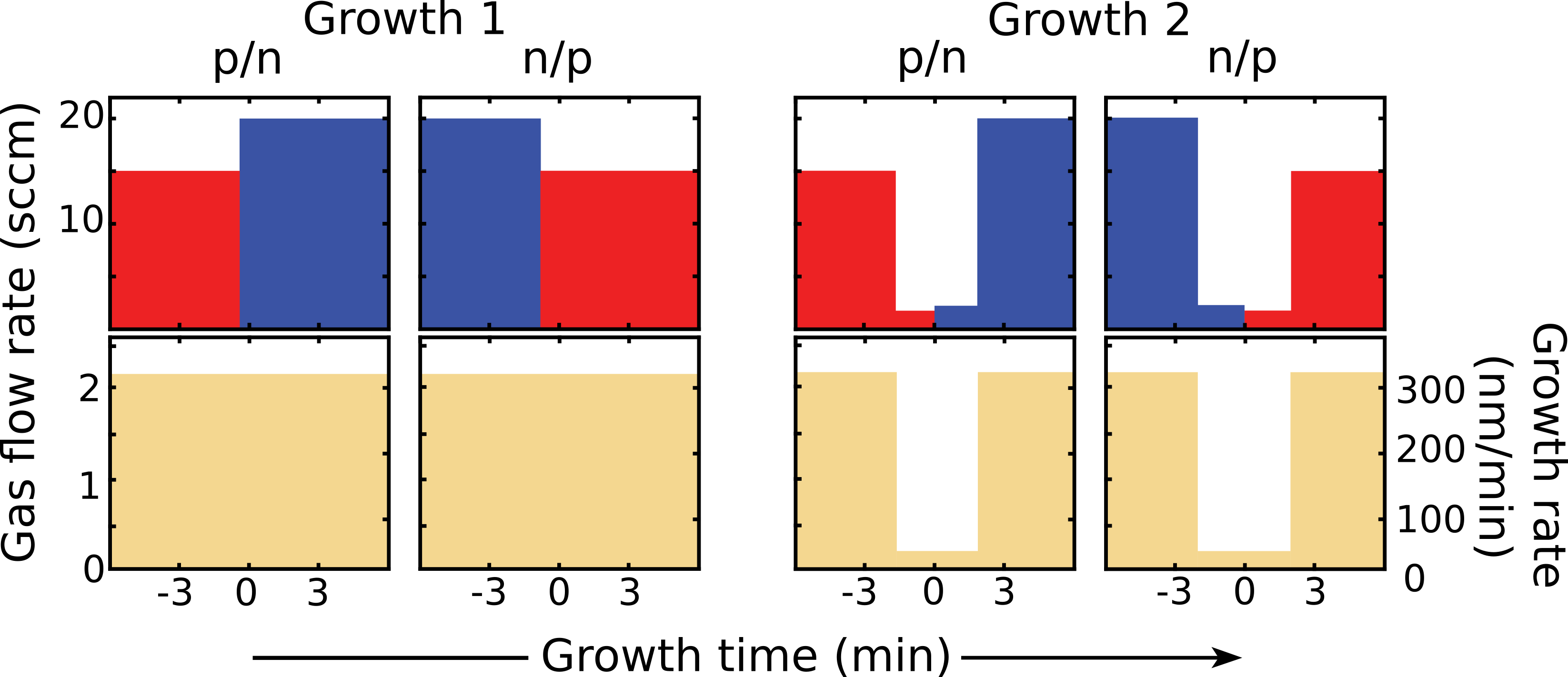}
\par\end{centering}
\caption{Gas flow switching schemes for each growth in both sequences \emph{p}/\emph{n}
and \emph{n}/\emph{p}. The B$_{2}$H$_{6}$, PH$_{3}$ and SiH$_{4}$
flows are plotted in red, blue and yellow, respectively. Note that
the growth direction is from left to right and that the gold catalyst
is always on the right throughout the document. We define time zero
as the point where the dopant precursors are switched. The SiH$_{4}$
flow rate is converted to a SiNW growth rate on the right-band axis.}
\end{figure}

Figure 2 shows a SE image via focused helium ion microscopy (HIM)
and a bright-field TEM image of representative NWs resulting from
Growth 1 in the \emph{p}/\emph{n} sequence. Images of SiNWs from other
growths did not vary from these results, except for the presence of
kinks nearby the junction, more commonly observed for NWs of Growth
2. In both Figure 2(a) and (b) the Au catalyst particle is visible
at the right end. The length of the SiNWs typically varied from 12
to 18 $\mu$m with the \emph{p-n} junction expected at 1.4 to 1.7
$\mu$m from the Au catalyst end. The HIM SE image shows strong contrast
at the lateral edges of each NW typical for this technique. There
is a reduction in intensity near holes in the support grid related
to transmission of the He ion beam through the sample. Although experimental
parameters such as magnification, ion beam energy and working distance
were chosen to maximize SE contrast, there was no sign of a \emph{p-n}
junction. The native oxide developed by these SiNWs is shown in the
TEM image detail in Figure 2(c). Similar to SEM imaging of \emph{p-n}
junctions, the likely presence of mid-gap interface states at the
oxide interface probably leveled any difference in the SE emission
rates between the\emph{ n} and \emph{p}-type segments, masking the
dopant transition. In Figure 2(b) the bright-field TEM contrast is
primarily due to variations in the intensity of diffraction. The thin
black line in the upper part of the NW corresponds to another much
smaller SiNW that adhered during TEM sample preparation. Other TEM
studies showing the development of a twin boundary bisecting some
of these NWs \citep{Sun} can be found in the Supplemental Information.

\begin{figure}[!htb]
\begin{centering}
\includegraphics{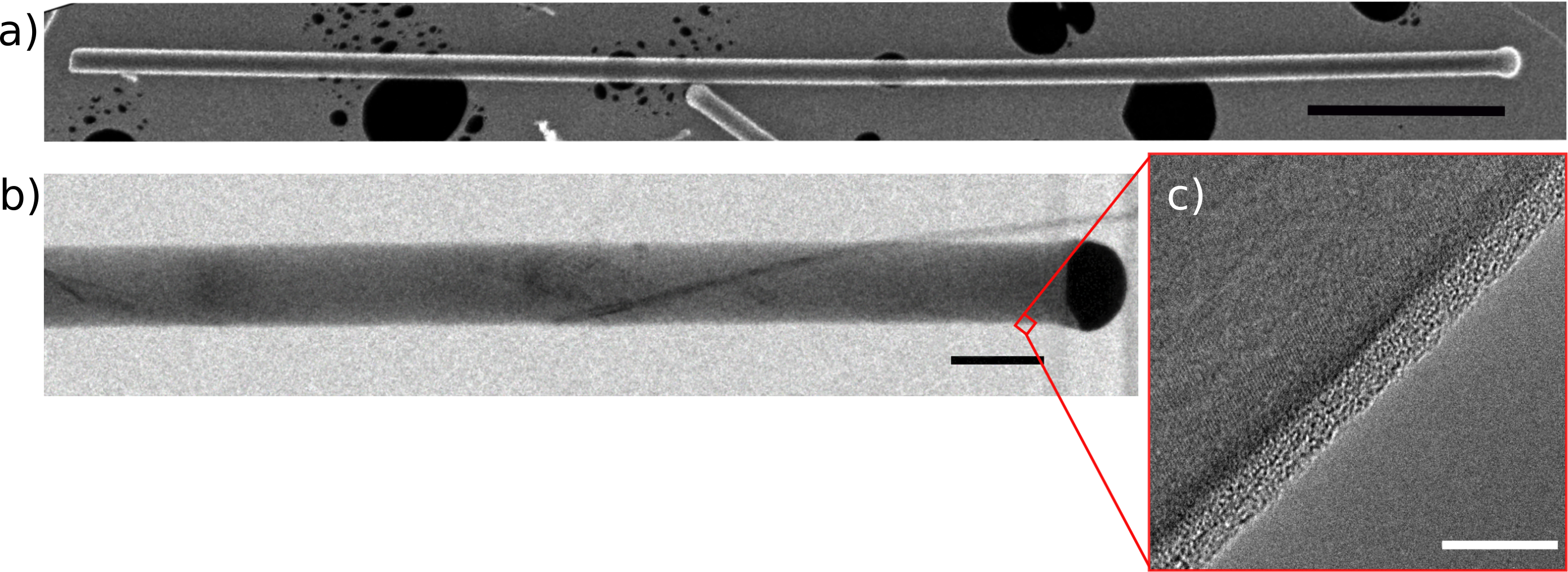}
\par\end{centering}
\caption{Imaging of SiNWs from Growth 1 in the p/n sequence. (a) Secondary
electron image via focused helium ion microscopy (note that the TEM
holey carbon support grid is visible in the background), (b) bright-field
TEM image, and (c) high magnification image of the region enclosed
by the area in (b). Scale bars are (a) 2 $\mu$m, (b) 200 nm and (c)
10 nm.}
\end{figure}

Figure 3 shows representative EH results for SiNWs resulting from
Growth 1, in both the \emph{p}/\emph{n} and\emph{ n}/\emph{p} growth
sequences. The contrast seen in the bright-field TEM image in Figure
3(c) is rather uniform and due entirely to differences in the degree
of diffraction from small angle bending and thickness variations,
except for small dark spots that accumulated at the junction area
indicative of a higher atomic number element (e.g. Au). Also, note
the smaller SiNW in Figure 3(c), that seems to have nucleated on a
particle located close by the junction. In Figure 3(d), we see strong
contrast from stacking fault defects nearby the junction area. This
type of defect developed independently of growth sequence or doping
type and was randomly distributed along most NWs. Figure 3(e) and
(f) show a reconstructed phase map with respect to the vacuum reference
in radians for the \emph{p}/\emph{n} and \emph{n}/\emph{p} sequences,
respectively. There is a clear decrease in phase from the\emph{ n}
to the\emph{ p}-type regions in both cases. Details of the transition
are shown in the average profile (Figure 3(g) and (h)) obtained from
the rectangular region enclosed by the blue rectangle in Figure 3(c)
and (h), respectively. The length of the depletion regions are $19\pm2$
and $12\pm1$ nm for the \emph{p}/\emph{n} and \emph{n}/\emph{p} growth
sequences, respectively. These values were extracted from fittings
to the potential using a linear step function and linear background,
details of which can be found in the Supplemental Information. We
have assumed that the NWs have a cylindrical geometry and that any
width variation is isotropic. Therefore, a measurement of the width
of the images was sufficient for calculating the potential difference
in the direction of the beam following Equation (1). Radial profiles
were taken from high-magnification bright-field TEM images at 10,
20, 50 and 100 nm from the junction. The V$_{bi}$ at the junction
was calculated to be $1.17\pm0.02$ and $0.86\pm0.01$ V for the \emph{p/n}
and \emph{n}/\emph{p} growth sequences, respectively.

\begin{figure}[H]
\begin{centering}
\includegraphics[scale=1.4]{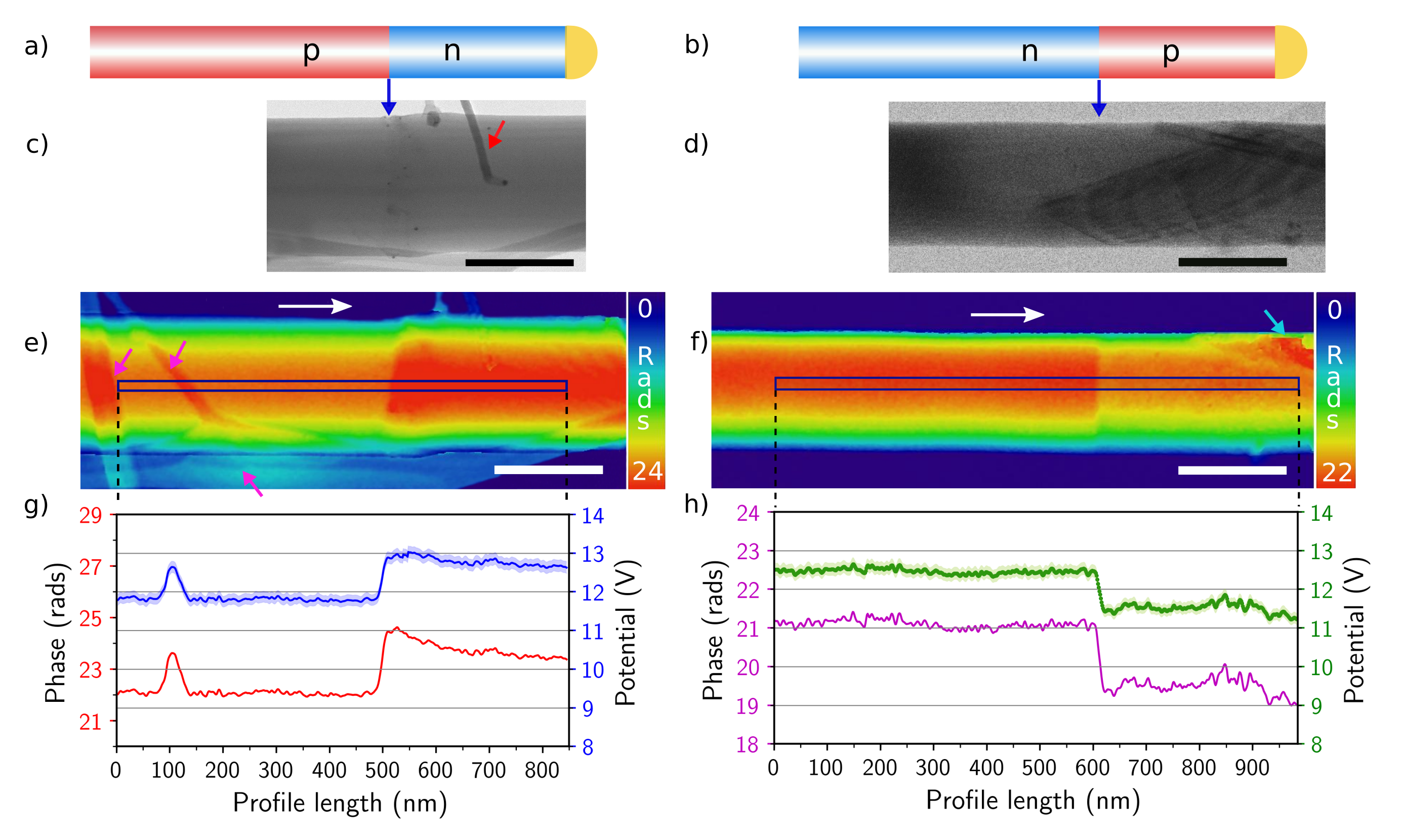}
\par\end{centering}
\caption{EH results for SiNWs from Growth 1. (a) \emph{p}/\emph{n} and (b)
\emph{n}/\emph{p} growth sequence schematics (not to scale), (c) and
(d) bright-field TEM images of the region nearby the \emph{p-n} junction,
(e) and (f) pseudo-colored phase images and (g) and (h) corresponding
1-dimensional phase and calculated potential profiles taken along
the center axis of the NWs as shown by the boxes regions in panels
(e) and (f). The blue and green error bands in (g) and (h) correspond
to the error added by the native oxide on the electron beam direction.
The blue arrows in (c) and (d) indicate the \emph{p-n}/\emph{n-p}
junction location. The white arrows in (e) and (f) indicate the NW
growth direction, while the fuchsia and cyan arrows indicate the presence
of the TEM carbon grid that supported the NWs, and a phase unwrapping
artifact, respectively. Note that in (e) and (f) the length of the
blue boxes correspond to the length of the profile that is plotted
in (g) and (h), and their width is used for averaging. Color bars
in (e) and (f) are in units of radians. Scale bars in (c)-(f) are
200 nm.}
\end{figure}

Figure 4 shows representative EH results for Growth 2. The contrast
of the bright-field TEM image for the \emph{p/n} growth sequence (Figure
4(c)) is also due to differences in the degree of diffraction from
small angle bending and thickness variations, while in the \emph{n}/\emph{p}
growth sequence (Figure 4(d)), we see strong contrast from defects
nearby the junction area. SE images and dopant delineation by etching
(not shown here) showed that the junction was located at $\sim$$1.3$
and $1.6$ $\mu$m from the Au catalyst for the\emph{ p}/\emph{n}
and \emph{n}/\emph{p} growth sequences, respectively. Figure 4(e)
and (f) show phase images in which there is a clear decrease in phase
from the \emph{n} to the \emph{p}-type regions in both cases. The\emph{
p}/\emph{n} (\emph{n}/\emph{p}) transition can be seen in detail by
averaging a line profile, shown in Figure 4(g) and (h), which was
obtained from the rectangular region enclosed by the blue rectangle
in Figure 4(e) and (f), respectively. These profiles show depletion
region widths of $10\pm1$ and $77\pm20$ nm for the\emph{ p}/\emph{n}
and \emph{n}/\emph{p} growth sequences, respectively. The \emph{p}/\emph{n}
growth sequence shows a $V_{bi}$ of $1.02\pm0.02$ V, while the \emph{n}/\emph{p}
shows a $V_{bi}$ of $0.8\pm0.3$ V. As for Growth 1, we have also
assumed that the NWs have cylindrical geometry and have used the projected
width to infer the specimen thickness in the beam direction for the
calculation of the potential.

\begin{figure}[H]
\begin{centering}
\includegraphics[scale=1.4]{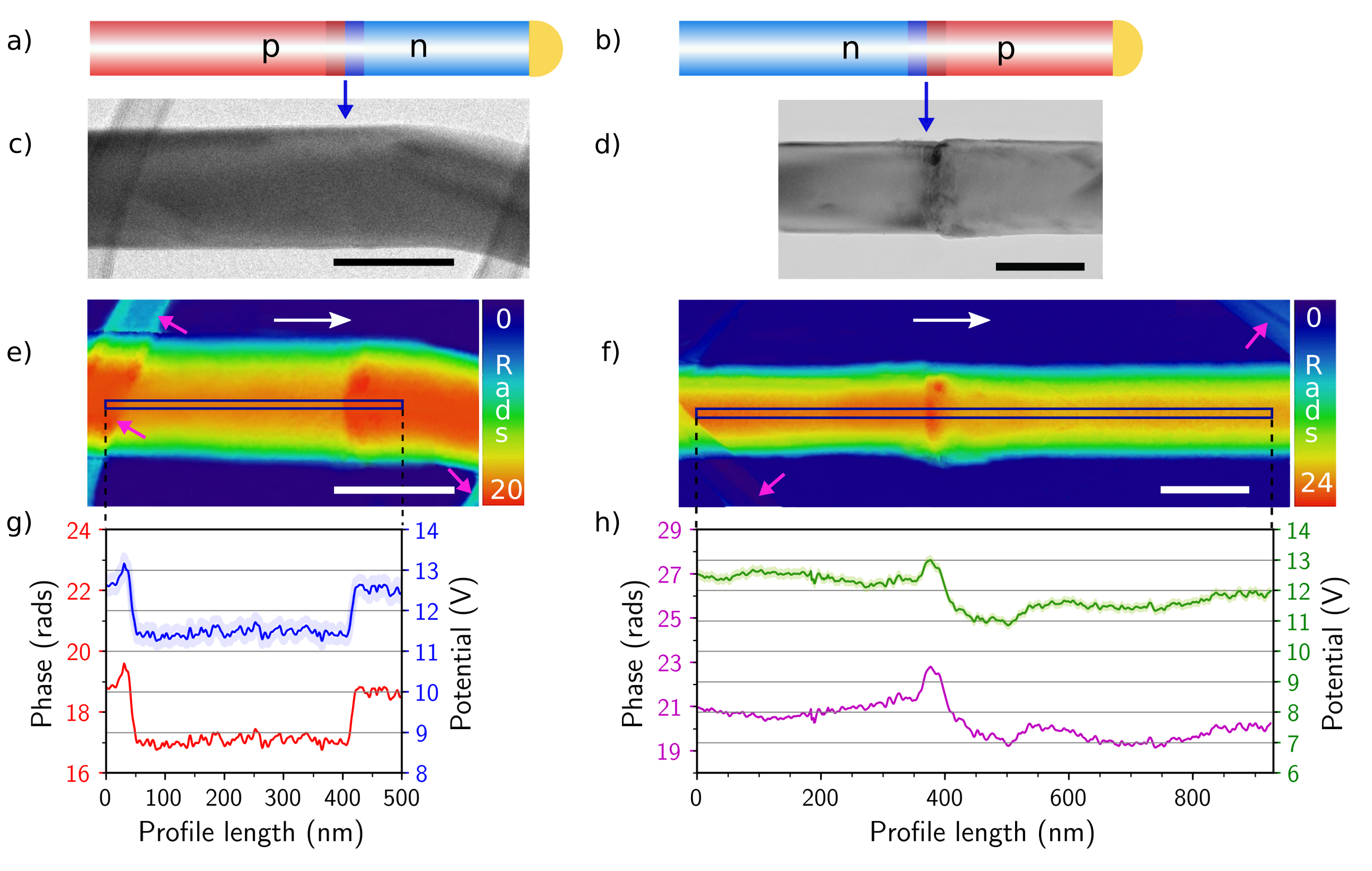}
\par\end{centering}
\caption{EH results for SiNWs from Growth 2. (a) \emph{p}/\emph{n} and (b)
\emph{n}/\emph{p} growth sequence schematics (not to scale). The colored
regions in the middle indicate the drop in pressure that differentiates
Growth 2 from Growth 1. (c) and (d) bright-field TEM images of the
region nearby the \emph{p-n} junction, (e) and (f) pseudo-colored
phase images and (g) and (h) corresponding 1-dimensional phase and
calculated potential profiles taken along the center axis of the NWs
as shown by the boxed regions in panels (e) and (f). The blue and
green error bands in (g) and (h) correspond to the error added by
the native oxide on the electron beam path. The blue arrows in (c)
and (d) indicate the \emph{p-n}/\emph{n-p} junction location. The
white arrows in (e) and (f) indicate the NW growth direction, while
the fuchsia arrows indicate the presence of the TEM carbon grid that
supports the NWs. Note that the length of the blue boxes in (e) and
(f) correspond to the length of the profile that is plotted in (g)
and (h), and their width is used for averaging. Color bars in (e)
and (f) are in units of radians. Scale bars in (c)-(f) are 200 nm.}
\end{figure}

\begin{figure}[H]
\begin{centering}
\includegraphics[scale=0.8]{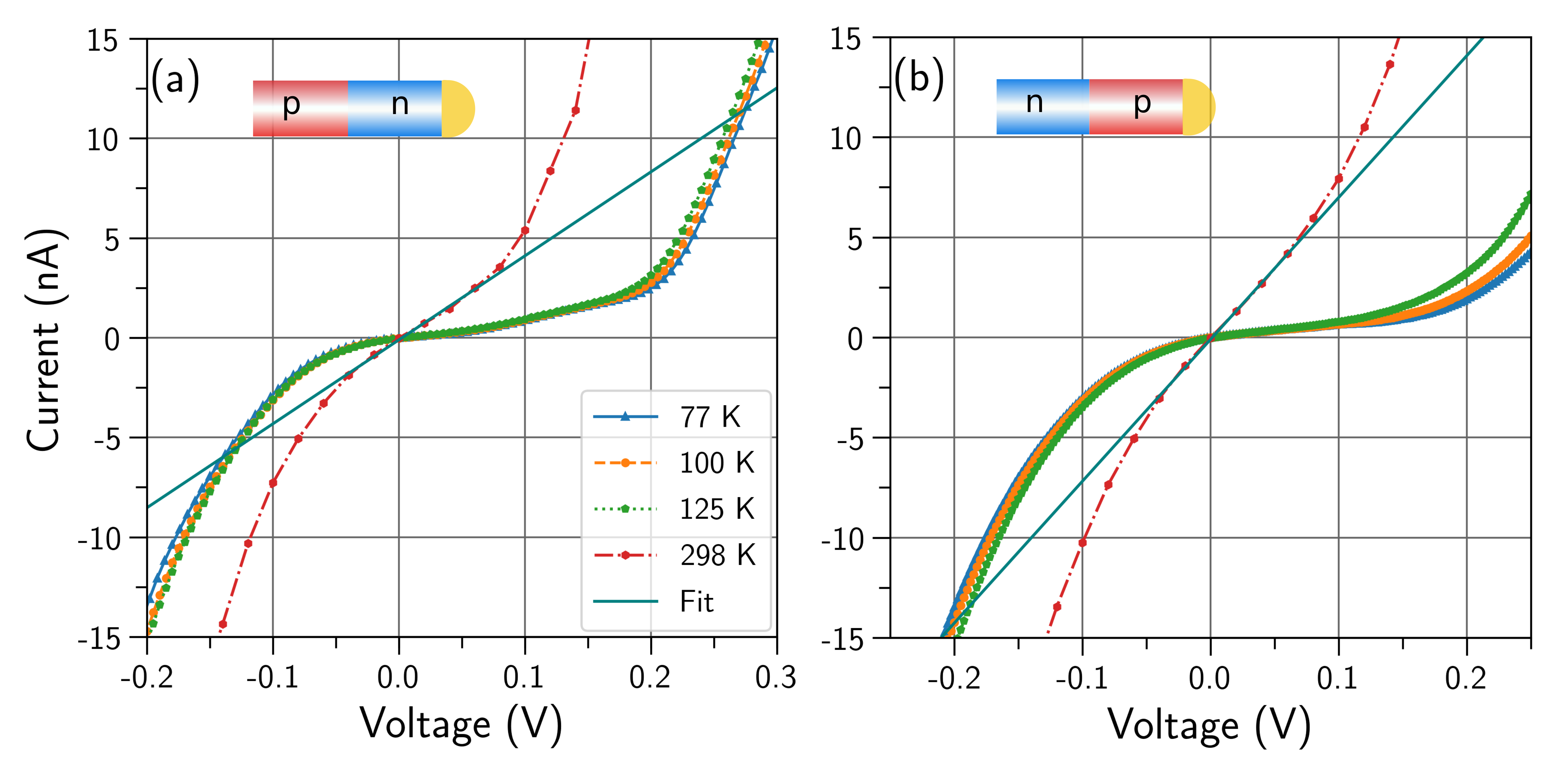}
\par\end{centering}
\caption{\emph{I}-\emph{V} sweep of two SiNWs from Growth 1 in the (a) \emph{p}/\emph{n}
and (b) \emph{n}/\emph{p} growth sequence at various temperatures.
Teal solid-lines show a linear fit to the low-bias data atroom-temperature,
yielding an effective resistance. Insets are the growth sequence schematics
(not to scale). The positive bias was applied to the \emph{p}-type
side in both cases. Teal solid-lines show a linear fit to the low-bias
data at room temperature.\label{fig:IV_growth1}}
\end{figure}

Figure~\ref{fig:IV_growth1} shows \emph{I}-\emph{V} sweeps for Growth
1 at various temperatures. We calculated the effective resistance,
which we define as the resistance times the cross-sectional area of
the NW. At room temperature (RT) SiNWs from Growth 1 behave as resistors
for biases less than $\sim$$50$ mV across the \emph{p}-\emph{n }junction,
yielding average effective resistance values of $6\pm6$ and $2.7\pm0.9$
m$\Omega\cdot\text{cm}^{2}$ for the \emph{p}/\emph{n} and \emph{n}/\emph{p
}sequences, respectively (see resistance data in Table S1). SiNWs
from Growth 2 (Figure S8) in the \emph{p}/\emph{n} growth sequence
exhibit similar behavior to those from Growth 1 at RT with an effective
resistance of $1.3\pm0.8$ $\Omega\cdot\text{cm}^{2}$. However, NWs
from Growth 2 in the \emph{n}/\emph{p} direction exhibit a substantially
lower resistance yielding an average effective resistance of $6\pm3$
$\mu\Omega\cdot\text{cm}^{2}.$ At $77$ K, most NWs exhibit an asymmetric
curvature in the forward bias direction with a change in slope near
$+0.2$ V. 

\section*{Discussion}

In the semiclassical treatment of \emph{p-n} junctions, the electric
potential $V\left(x\right)$ satisfies Poisson's equation \citep{AshcroftMermin1}:

\begin{equation}
-\frac{d^{2}V}{dx^{2}}=\frac{\rho\left(x\right)}{\epsilon},\quad(\text{SI units}),\label{eq:Poisson's equation in SI units - 1}
\end{equation}

\noindent where $\epsilon=\epsilon_{r}\epsilon_{0}$, is the permitivity
of the semiconductor prior to doping, $\epsilon_{r}$ and $\epsilon_{0}$
are the relative and vacuum permitivities respectively, and $\rho\left(x\right)$
is the charge distribution. For sufficiently low doping, Eq.~\eqref{eq:Poisson's equation in SI units - 1}
can be solved assuming Boltzmann statistics, within the so-called
full depletion approximation, where one obtains an analytical solution
for the width of the depletion region \citep{Streetman}. However,
estimations for acceptor ($N_{A}$) and donor ($N_{D}$) impurity
concentrations from uniformly doped \emph{p} and \emph{n}-type SiNWs
grown under similar conditions, are $\sim$$4.5\times10^{19}$ and
$\sim$$1\times10^{20}\text{cm}{}^{-3}$, respectively, which are
degenerate doping levels. And, our measured $V_{bi}$, $\sim$$1.0$
eV, compares with the Si bandgap ($\sim$$1.1$ eV). Therefore, we
also solved Eq.~\eqref{eq:Poisson's equation in SI units - 1} numerically
using Fermi-Dirac statistics. For the aforementioned $N_{A}$ and
$N_{D}$ doping concentrations, and assuming an abrupt doping profile,
we obtained a built-in potential of 1.173 V and a depletion region
width of 4.7 nm. Details of this calculation can be found in the Supplemental
Information.

\begin{figure}[H]
\begin{centering}
\includegraphics[scale=0.45]{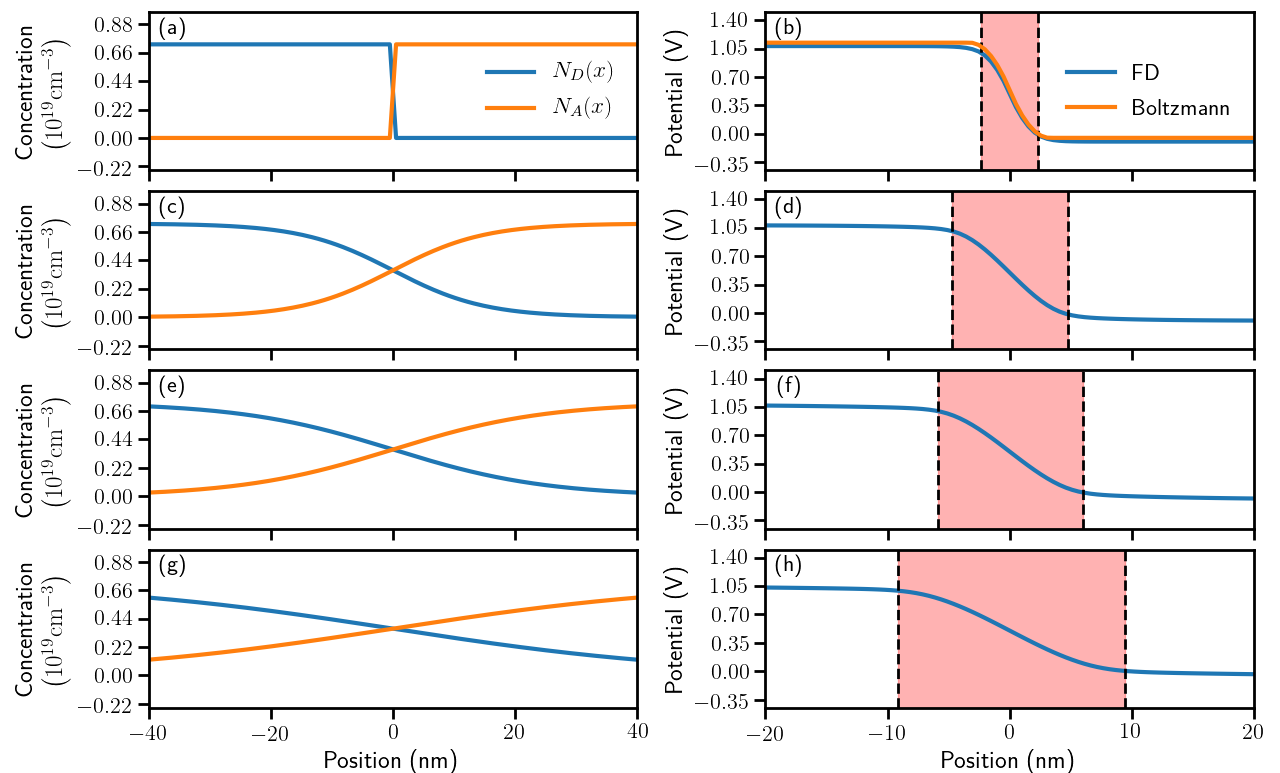}
\par\end{centering}
\caption{Doping concentration and associated junction potential as a function
of position at $T=300$ K for $N_{A}=N_{D}=7.25\times10^{19}$ cm$^{-3}$
using Fermi-Dirac statistics: (a), (b) an abrupt doping profile yielding
a depletion region width of 4.7 nm; graded doping profiles modeled
by a hyperbolic tangent curves yielding depletion region widths of
(c), (d) 9.5; (e), (f) 11.9 and (g), (h) 18.6 nm. The pink shaded
areas show the depletion region width measured using the same procedure
as was applied to the experimental data. Note that we include the
Boltzmann statistics result for an abrupt doping profile yielding
a depletion width of 6.4 nm in (b). \label{fig:FD doping and potential}}
\end{figure}

Our measured depletion regions were larger than the prediction for
abrupt junctions. Thus, it is interesting to model our junctions with
a graded doping profile. Figure~\ref{fig:FD doping and potential}
compares the doping concentration profiles and depletion region widths
calculated using both Boltzmann and Fermi-Dirac statistics, assuming
maximum donor and acceptor impurity concentrations of $N_{A}=N_{D}=7.25\times10^{19}$
cm$^{-3}$ (i.e. the average of the experimental values of $N_{A}$
and $N_{D}$) and hyperbolic tangents for the graded doping transition
(see Supplemental Information). Fits for these profiles were carried
out using the same approach that was applied to the experimental data
for meaningful comparison.

\begin{figure}[H]
\begin{centering}
\includegraphics[scale=0.5]{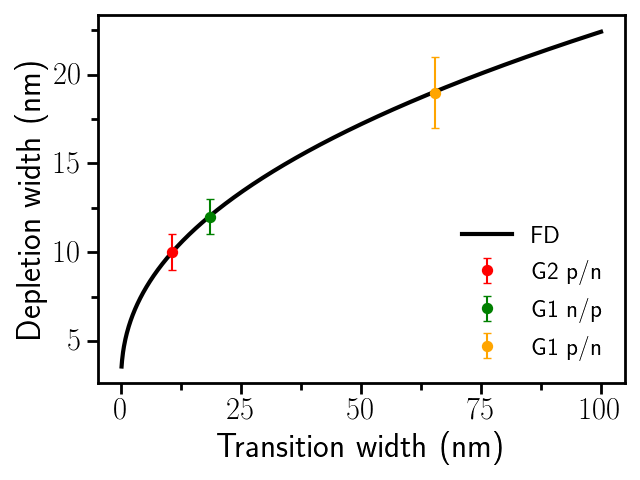}
\par\end{centering}
\caption{Depletion region width versus doping transition width calculated using
Fermi-Dirac statistics and graded doping profiles modeled via hyperbolic
tangents at T = 300 K. The data points with error bars correspond
to the depletion region widths measured via EH, except for Growth
2 in the \emph{n}/\emph{p} growth sequence, which is out of range.
\label{fig:depletion width and transition width}}
\end{figure}

Figure \ref{fig:FD doping and potential} plots the doping profiles
and the respective electric potential $V\left(x\right)$, as a function
of the doping transition width ($w$), assuming $w=0$, 10, 20 and
60 nm, giving rise to depletion widths of $\text{\ensuremath{4.7}}$,
$9.5$, $11.9$ and $18.6$ nm, respectively. The calculated depletion
region width versus the doping transition width is plotted in Figure~\ref{fig:depletion width and transition width}.
A graded transition increases the width of the depletion region, as
expected, and the values obtained are comparable to the measurements
via EH.

\begin{figure}[H]
\begin{centering}
\includegraphics[scale=1.5]{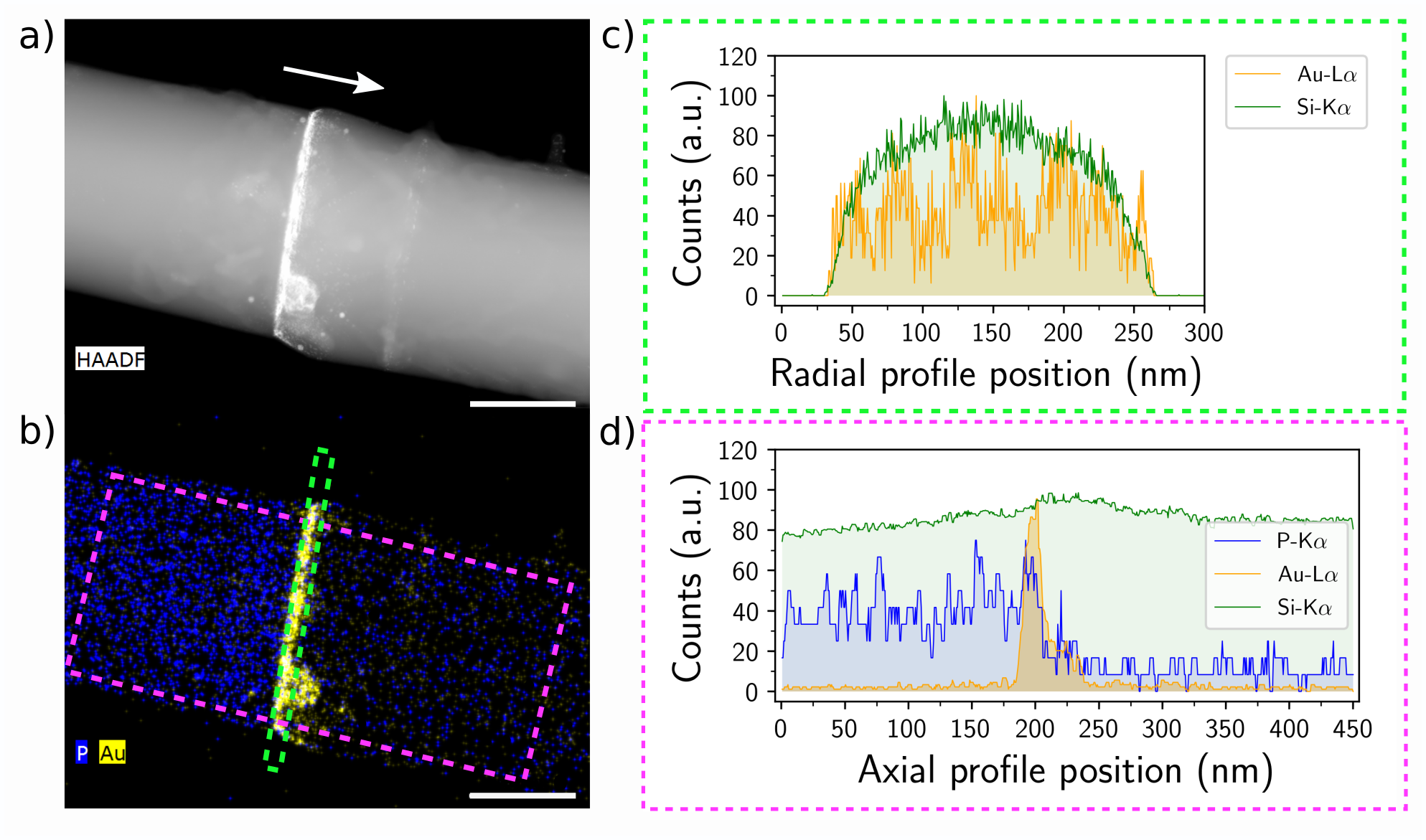}
\par\end{centering}
\caption{HAADF STEM analysis of a SiNW from Growth 2 in the \emph{n}/\emph{p}
sequence. (a) image of the region nearby the \emph{p-n} junction,
(b) EDS map of the same region as image (a), the Au and P signals
are depicted in yellow and blue, respectively. (c) Au and Si EDS counts
as a function of radial profile position taken along the green box
in (b). (c) Si, Au and P EDS counts as a function of axial profile
position taken along the fuchsia box in (b). The white arrow in (a)
indicates the NW growth direction. Scale bars in (a) and (b) are 100
nm.\label{fig:EDX}}
\end{figure}

All measured depletion regions were less than 20 nm with the exception
of the \emph{n}/\emph{p} sequence of Growth 2, which showed a long
depletion region ($77\pm20$ nm) and highly defective junction. Existence
of these junctions, including an irregular one corresponding to this
growth and sequence, was confirmed via EBIC measurements (see Supplemental
Information). In our experiments (Growth 2), we slowed down the growth
rate nearby the junction to avoid the reservoir effect by allowing
P or B to evaporate from the liquid catalyst. A high-angle annular
dark-field (HAADF) STEM image as well as EDS maps and profiles of
the defective \emph{n}/\emph{p} junction of Growth 2 are shown in
Figure~\ref{fig:EDX}. Accumulation of Au is confirmed by the high-contrast
regions in the HAADF image and by the EDS map of Au in Figure~\ref{fig:EDX}(b).
Figure~\ref{fig:EDX}(c) and (d) show line profiles comparing the
Si and Au EDS signals along the green and fuchsia boxes in Figure
~\ref{fig:EDX}(b). A simple median filter was applied to the EDS
data as a method of noise reduction \citep{Stone}.

Au deposited at the junction location (Growth 2, \emph{n}/\emph{p}
sequence) according to Figure 8; however, SiNW growth was not hampered,
as NWs reached their intended lengths ($\sim$$20$ $\mu$m). Higher
phase shifts (darker red coloration in Figure 4(f)) in the phase image
as well as the profile peak (Figures 4(f) and (h)) are also a signature
of Au accumulation at the junction. A calculation of the amount of
Au in the direction of the beam for this profile yielded a maximum
thickness of 9 nm (see Supplemental Information). Au is known to act
as a surface passivator stabilizing epitaxial growth of <111> SiNWs
\citep{Pan,Kim,Schmid,Dailey}, and whenever this passivation is inhibited,
SiNWs kink towards the more thermodynamically stable <112> direction.
Au decoration is also known to happen randomly in highly-doped NW
segments, dependent on  pressure, growth rate, changes in the Au catalyst
liquid volume and surface chemistry. Pressure and growth values exceeding
100 mTorr and 150 nm/min, respectively, are necessary to minimize
this effect \citep{Kim,Dailey}.

Accumulation of Au at the junction interface is correlated with changes
in the partial pressure of the precursors as described in Figure 1.
If the change in SiH$_{4}$ partial pressure alone was the cause,
one would expect large Au deposits in the opposite growth sequence
(\emph{p}/\emph{n}) as well, since the SiH$_{4}$ partial pressure
changes by the same magnitude. However, that was not the case. SiNWs
resulting from Growth 2 in the \emph{p}/\emph{n} sequence showed negligible
accumulation of Au. The most likely explanation is that the order
in which one switches the gas precursors (PH$_{3}$/B$_{2}$H$_{6}$)
influences the Au segregation. The EDS map and profile in Figure~\ref{fig:EDX}(b)
and (d), respectively, for growth in the \emph{n}/\emph{p} sequence
show a $\sim$$40$ nm transition of the P (\emph{n}-type) dopant
at the \emph{p-n} interface. Relatively short doping transitions are
expected since we followed growth conditions established in previous
studies that lead to sharp P transitions\citep{Christesen}. The B
dopant profile was not detected via EDS or electron energy loss spectroscopy
due to a peak overlap produced by the Au and B signals. Figure~\ref{fig:EDX}(d)
also shows a linear increase in the Si signal up to the junction region,
where the Au has accumulated, while it remains constant in the \emph{n-}type
segment of the NW. This increase is likely due to changes in the AuSi
catalyst contact angle during growth.

We did not observe charging induced by the electron beam in these
NWs, therefore we can assume that the MIP values extracted from the
calculated voltage profiles should occur at the midpoint of the voltage
difference. The average values for both growths is $12.1\pm0.1$ V.
Reported MIP values for Si are $12.5\pm0.7$, $9.3\pm0.1$ V, and
$11.5\pm0.5$ V experimentally determined through EH \citep{Kruse1,Wu}.
Notice that our average MIP value is within the error of two of the
previously reported values. 

We obtained V$_{bi}$ values that are comparable to the Si bandgap
at RT in the \emph{p}/\emph{n} growth sequence for both types of growth.
This was unexpected, since it has also been observed that measured
V$_{bi}$ values via EH can be up to 55\% lower than expected \citep{Wolf2013}.
Although these differences are known to increase for materials with
larger bandgaps \citep{Yazdi} and lower doping levels \citep{Houben}.
Interaction of the electron beam with the semiconducting NW generates
currents due to SE emission and electron-hole pair generation, which
can forward bias NWs decreasing the expected $V_{bi}$\citep{Cordoba}.
We attribute this apparent beam insensitivity to high dopant levels
that decrease the NW carrier mobility and diffusion lengths, decreasing
any beam induced charge accumulation. The long NW lengths created
multiple contact points to the carbon support grid, aiding the neutralization
of any shunt resistance developed while recording the data. However,
we consistently obtained $\sim$0.2 - 0.3 V lower $V_{bi}$ values
for SiNWs in the \emph{n}/\emph{p} growth sequence. This could be
attributed to residual PH$_{3}$ in the reactor that causes compensation
doping in the \emph{p}-type segment. These values could be low enough
to be out of the EDS detection limit and thus possibly at the noise
level in Figure~\ref{fig:EDX}.

The characteristic NDR observed in tunnel junction \emph{I}-\emph{V}
measurements was hidden likely due to excess currents originating
from trap-assisted (TA) tunneling through mid-bandgap trap states\citep{Sah}.
Note that the indirect bandgap of Si tends to favor TA tunneling because
of the momentum mismatch for direct band-to-band (BB) tunneling. The
TA tunneling rate is also substantially higher than BB tunneling except
in very high electric fields \citep{Schenk}. Unlike band-to-band
tunneling, trap-assisted tunneling is temperature-dependent and cooling
down the device can increase the peak-to-valley current ratio and
reveal NDR behavior \citep{SchmidH}. The change in slope observed
near $+0.2$ V (Figure~\ref{fig:IV_growth1}) in low temperature
\emph{I}-\emph{V} curves resemble those previously seen in Au-doped
planar Si tunnel junctions \citep{Sah}, suggesting that our tunnel
junctions are conducting via the same mechanism. Since we have little
control over the gold defect density in the NWs during our VLS growth
process, the resistance of the NWs can vary noticeably between NWs
from the same synthesis. However, NDR, a hallmark of tunneling behavior,
was observed for one NW (Figure S8) likely because it had an unusually
low Au defect density at the junction, but also because it was grown
in the Growth 2 \emph{p}/\emph{n} sequence, which exhibited the shortest
depletion width of all conditions studied. At the opposite extreme,
when Au concentration is very high, such as those observed via EDS
in Growth 2\emph{ n}/\emph{p} sequence, the junctions are electrically
shorted, and their average resistivity ($12\pm6$ m$\Omega\cdot\text{cm}$)
is consistent with reported values for heavily-doped \emph{n}- and
\emph{p}-type bulk Si \citep{Masetti}. The other growth sequences
exhibit higher resistivity values (Table S1) as expected, dominated
by the junction resistance.

\section*{Conclusions}

We have used electron holography (EH) to characterize depletion regions
of heavily-doped, VLS-grown SiNWs. We compared the effects of growth
conditions and sequence order: \emph{p}/\emph{n} and \emph{n}/\emph{p}
with the aim of optimizing tunneling junctions. Keeping the partial
pressure of the precursors constant (Growth 1) yielded narrow depletion
regions of $19\pm2$ nm and $12\pm1$ nm for \emph{p}/\emph{n} and
\emph{n}/\emph{p}, respectively. Slowing the growth rate at the junction
region (Growth 2) increased the junction abruptness, decreasing the
depletion width to $10\pm1$ nm in the case of the \emph{p} to \emph{n}
switching. However, slower switching also created more instabilities
in the Au catalyst leading to more kinking in the majority of the
NWs. Slower growth in the \emph{n} to \emph{p} sequence substantially
increased the degree of segregation of Au particles from the catalyst
seed. Measured V$_{bi}$ values were higher and comparable to the
Si bandgap for the \emph{p}/\emph{n} sequence ($1.17\pm0.02$ and
$1.02\pm0.02$ V) than for the \emph{n}/\emph{p }sequence ($0.86\pm0.01$
and $0.8\pm0.3$ V) in both types of growth, potentially due to PH$_{3}$
doping compensation in the \emph{p}-type segment of the \emph{n}/\emph{p}
sequence.

Overall, the results have confirmed the existence of electrically
abrupt, degenerately-doped axial SiNW junctions that exhibit tunnel-junction
behavior in both sequences (\emph{p}/\emph{n} and \emph{n}/\emph{p}).
The width of the depletion regions could be as short as $10\pm1$
nm using optimal growth rates and sequences. This is the first essential
step in the creation of more complex solar energy devices and possibly
other technologies that require tunnel junctions, such as tunneling
field-effect transistors and multijunction solar cells.

\section*{Methods}

\subsection*{Nanowire synthesis}

SiNWs were synthesized in a home-built chemical vapor deposition (CVD)
system \citep{Christesen,Hill2017,Pinion,Kim}. Au catalysts of $\sim$$200$
nm diameter were immobilized on SiO$_{2}$ wafers. This thermally-grown
SiO$_{2}$ (300 nm) in the wafer made the nucleation and growth happen
preferably in the <112> growth direction. Silane (SiH$_{4}$) was
the Si source, diborane (B$_{2}$H$_{6}$, 1000 ppm in H$_{2}$) and
phosphine (PH$_{3}$, 1000 ppm in H$_{2}$) served as \emph{p} and
n-type dopant precursors, respectively. Hydrochloric acid (HCl ) was
used for chlorination and H$_{2}$ as the carrier gas. Growth was
carried out at 510$^{\circ}$C, total pressure of 20 Torr and a HCl:SiH$_{4}$
ratio of 2:1 for both types of growth.

\subsubsection*{Growth 1}

For the \emph{p}/\emph{n} sequence, we first grew 10 - 15 $\mu$m
highly-doped\emph{ p}-type at a growth rate of 320 nm/min and abruptly
switched the dopant precursor to\emph{ n}-type with a growth rate
of 260 nm/min, until the end of the NW growth (1.4 $\mu$m). For the\emph{
n}/\emph{p} sequence we first grew 5 $\mu$m \emph{p}-type followed
by 12 $\mu$m of highly \emph{n}-doped segment at a growth rate of
280 nm/min until abruptly switching to \emph{p}-type with grow rate
of 280 nm/min for five minutes (1.6 $\mu$m) until the end of the
growth.

\subsubsection*{Growth 2}

For the \emph{p}/\emph{n} sequence, we first grew 10 - 15 $\mu$m
of highly \emph{p}-doped segment at a growth rate of 320 nm/min, then,
we dropped the growth rate to 32 nm/min for two minutes until abruptly
stopping the flow of the diborane precursor. At the same time the
phosphane precursor was turned on with a growth rate of 28 nm/min.
Finally, after 2 mins ($\sim$$56$ nm) the growth rate was increased
back to 280 nm/min. For the opposite sequence (\emph{n}/\emph{p}),
nucleation started with 3 $\mu$m of undoped Si, followed by 12 $\mu$m
of\emph{ n}-doped Si at a growth rate of 280 nm/min. Then the growth
rate was decreased to 28 nm/min for two minutes until abruptly stopped.
The \emph{p}-type precursor was started immediately after at a growth
rate of 32 nm/min for two minutes. Finally, the growth rate was increased
back to 320 nm/min. The Si precursor to dopant precursor ratios (SiH$_{4}$:PH$_{3}$,
SiH$_{4}$:B$_{2}$H$_{6}$) remained constant in both cases.

\subsection*{TEM sample preparation and imaging}

TEM sample preparation was carried out by mechanical abrasion of NWs
directly onto lacey carbon support grids, to avoid artifacts or NW
modification created by solvents. EH data was collected using electrons
generated by a field-emission gun, operated at an acceleration voltage
of 200 kV in a scanning transmission electron microscope (STEM) with
point to point resolution of 0.23 nm (FEI Tecnai G2) using the objective
lens for TEM imaging and a Lorentz lens for EH, which gives a wider
field of view. The exposure time was 1.5 s and the biprism voltage
140 V. The reference hologram was obtained 800 nm from the NW axis
for a magnification of 33 kX, following the optimum position for an
empty hologram procedure \citep{Wolf1,Kou}. Higher magnification
images were acquired on a Hitachi HF-3300 TEM, HF-3300V scanning transmission
electron holography microscope (STEHM) cold-field-emmision electron
source, operated at an acceleration voltage of 300 kV which can achieve
both a STEM electron probe size and TEM spatial resolution approaching
45 pm. Finally, STEM and energy dispersive x-ray spectroscopy maps
were obtained using a STEM (FEI Osiris) operated at 200 keV with a
STEM resolution of 0.16 nm.

\subsection*{Current-voltage measurements}

SiNWs were mechanically transferred from the growth substrate onto
Si/SiO$_{2}$/SiN$_{4}$ wafers. A photo-resist stack (MMA EL9 and
PMMA A2, Microchem) was spun on top of the NWs and contacts to single
NWs were patterned via electron beam lithography (Nanometer Pattern
Generation System). The native oxide was etched from the patterned
area using buffered hydrofluoric acid (Transene) and metal electrodes
(3 nm Ti, 300 nm Pd) were deposited with electron beam evaporation
(KJ Lesker PVD 75). Each NW was contacted by four electrodes to ensure
that the contacts were ohmic. Current-voltage measurements at ambient
conditions were taken with a Keithley 2636A SourceMeter connected
to a pair of micropositioners (Signatone, S-725) and probe tips (SE-TL,
tungsten). For low temperature measurements, the NWs were tested in
a cryogenic probe station (Lakeshore PS-100) under vacuum and cooled
with liquid nitrogen.

\ack{}{}

We are grateful for partial funding from the Canadian Natural Science
and Engineering Research Council (NSERC), the Canadian Foundation
for Innovation, the BC Knowledge Development Fund, and SFU 4D Labs. 

T.S.T. and J.F.C. were supported by the University of North Carolina
Energy Frontier Research Center (EFRC), \textquotedblleft Alliance
for Molecular Photoelectrode Design\textquotedblright{} (AMPED), funded
by the U.S. Department of Energy, Office of Science, Office of Basic
Energy Sciences, under Award DE-SC0001011. This work made use of instrumentation
at the Chapel Hill Analytical and Nanofabrication Laboratory (CHANL),
a member of the North Carolina Research Triangle Nanotechnology Network
(RTNN), which is supported by the National Science Foundation (ECCS-1542015)
as part of the National Nanotechnology Coordinated Infrastructure
(NNCI).

\bibliographystyle{unsrt}
\addcontentsline{toc}{section}{\refname}\bibliography{bibfile}

\end{document}